\documentstyle[12pt]{article}

\begin{document}
\title{ Two-dimensional gases of generalized statistics
\\ in a uniform magnetic field}
\author{Choon-Lin Ho and Man-Jui Liao}
\date{\small Department of Physics, Tamkang University, Tamsui,
      Taiwan 25137, R.O.C.}

\maketitle

\begin{abstract}

We study the low temperature properties of two-dimensional ideal
gases of generalized statistics in a uniform magnetic field.  The generalized
statistics considered here are the parafermion statistics and the exclusion
statistics.
Similarity in the behaviours of the parafermion gas of finite order $p$
and the gas with exclusion coefficient $g=1/p$ at very low temperatures is
noted. These two systems become exactly equivalent at $T=0$.
Qumtum Hall effect with these particles as charge carriers is briefly
discussed.

\end{abstract}

\newpage

\section{Introduction}

The fractional quantum Hall effect (FQHE)  and the theory of anyon
superconductivity have aroused increasing interest in generalized statistics.
These two systems involves particles called anyons, which
obey fractional exchange statistics \cite{FS}.
The concept of anyons is essentially two-dimensional (2D).  More recently,
Haldane \cite{Haldane} proposed a different definition of fractional
statistics which
is independent of space dimensionality.   It involves a generalisation of the
Pauli exclusion principle, and is thus commonly termed the exclusion
statistics. Important examples of particles obeying the exclusion statistics
are the quasi-particles
in the FQHE (\cite{Haldane}-\cite{Shirai}).

Historically, the first proposal of a generalized statistics is the
parafermion statistics of finite order $p$ ($p$ being
positive integers) (\cite{GG}, \cite{Gb}), in which
at most $p$ particles are allowed to occupy a single quantum state.
This is the first generalisation of the Pauli exclusion principle.
It was speculated that quarks might be parafermions of order $p=3$
\cite{Gb}.
The statistical distribution ${\bar n}_i$ of an ideal parafermion gas of order
$p=1,2,\ldots$
with chemical potential $\mu$ and temperature $T$ is given by \cite{AVT}
\begin{eqnarray}
{\bar n}_i=\frac{1}{e^{\left(E_i-\mu\right)/k_B T}-1}
   -\frac{p+1}{e^{\left(p+1\right)\left(E_i-\mu\right)/k_B T}-1}~.
\label{n-p}
\end{eqnarray}
where $E_i$ is the energy spectrum of a single particle.
One notes that (\ref{n-p}) reduces to the Fermi-Dirac and Bose-Einstein
statistics for $p=1$ and $p\to\infty$, respectively.
It can be easily checked that at $T=0$:
\begin{eqnarray}
{\bar n}_i=\cases{p~, & if $E_i < \mu$~;\cr
           0  ~, & if $E_i > \mu$~.\cr}
\label{n-p-2}
\end{eqnarray}
Thus for finite $p$ the distribution (\ref{n-p}) exhibits a ``Fermi" surface.

In his alternative generalisation  of the Pauli exclusion principle,
Haldane \cite{Haldane} proposed that the change of available one-particle
Hilbert
space dimension $\Delta d$ to be related linearly to the change of the number
of particles $\Delta N$ by $\Delta d=-g\Delta N$, where $g$, called the
exclusion coefficient, is a measure of the degree of exclusion of the system.
Following \cite{NayWil}, we shall call particles with such property
$g$-ons. Thus $g$-ons with $g=0$ and $g=1$ correspond to bosons and
fermions, respectively.
Quasiholes of the Laughlin quantum Hall states with filling factors
$\nu=1/m$ have exclusion coefficents $g=1/m$ (\cite{Haldane}, \cite{JohnCan}).
The statistical distribution ${\bar n}_i$ of an ideal $g$-gon gas
is given by \cite{Wu} (see also \cite{Raja}):
\begin{eqnarray}
{\bar n}_i=\frac{1}{\omega_i(\zeta_i)+g}~,
\label{n-g}
\end{eqnarray}
where the function $\omega_i (\zeta_i)$ is solved from the functional equation
\begin{eqnarray}
\omega_i(\zeta_i)^g \left[1+\omega_i(\zeta_i)\right]^{1-g}=\zeta_i\equiv
e^{(E_i-\mu)/k_B T}~.
\label{w}
\end{eqnarray}
The distibution (\ref{n-g}) also exhibits a ``Fermi" surface at $T=0$:
\begin{eqnarray}
{\bar n}_i=\cases{1/g~, & if $E_i < \mu$~;\cr
           0  ~, & if $E_i > \mu$~.\cr}
\label{n-g-2}
\end{eqnarray}
From this one notices that the exclusion statistics with $g=1/p$ ($p$ being
positive integers) is equivalent to the parafermion statistics of order $p$ at
$T=0$. In fact,  they are almost completely identical at sufficiently low
temperatures.  This is evident from Fig.~1, which shows plots of ${\bar n}_i$
of both statistics as a function of $E_i/\mu$ at $k_B T=\mu/15$.
It is of interest to note that the quarks and the quasiholes of the Laughlin
$\nu=1/3$ quantum Hall state may obey the same statistics at low temperatures
(and in some cases, may even have the same magnitude of fractional charge
$e/3$).

Statistical properties of ideal parafermion gas were studied in \cite{AVT},
and those of ideal $g$-on gas in [11, 13, 14].
In this paper , we would like to investigate the behaviours of the ideal gas
of both types of particles in a uniform magnetic field at low temperatures.
We shall assume minimal coupling of these particles with the magnetic field.

\section{2D parafermion gas}

Let us first consider an ideal 2D
parafermion gas of finite order $p$ in a uniform
magnetic field $B$ pointing along the $z$-axis.
The energy spectrum of the 2D parafermion is
\begin{eqnarray}
E_n={\hbar e^\ast B\over m^\ast c}\left(n+{1\over 2}\right)~,
~~~~n=0,1,2,\ldots
\label{e2}
\end{eqnarray}
where $n$ is the number of Landau level, $m^\ast$ and $e^\ast$ are the
effective mass and the magnitude of the effective charge, respectively, of the
parafermion.

Thermodynamical properties of the ideal gases are
derivable
from the thermodynamical potential $\Omega(\mu,\,B,\,T)$, considered
as a function
of the chemical potential $\mu$, the external field $B$,
and the gas temperature $T$.  We only consider particles with positive chemical
potential ($\mu>0$) in this paper. The thermodynamical potential of the 2D
parafermion gas is given by :
\begin{eqnarray}
\Omega (\mu,\,B,\,T)=-k_B T n_2~ \sum_n\left[
\ln\left(1-e^{\left(\mu-E_n\right)\left(p+1\right)/k_B T}\right)
-\ln\left(1-e^{\left(\mu-E_n\right)/k_B T}\right)\right]~.
\label{e3}
\end{eqnarray}
Here $L_x,~L_y$ are the lengths of the 2D space,
$k_B$ is the Boltzmann constant,
and $n_2\equiv e^\ast BL_xL_y/ hc$ ($h=2\pi \hbar$)
is the degeneracy of the Landau levels.

Evaluation of the $\Omega$ is easily carried out by the Mellin transformation
\cite{Kha} with respect to the variable $\exp(a\mu/k_B T)$ $(a=1,p+1)$,
resulting in the following expression
\begin{eqnarray}
  \Omega(\mu,\,\kappa,\,T)=- \frac{1}{2}
   k_B T n_2\sum{\rm Res}\left[{\pi\cos(\pi s)\over s
  \sin(\pi s)}
{\exp(sa\mu/k_B T)\over\sinh(sa\kappa/2k_B T)}
  \right]^{a=p+1}_{a=1},
\label{Omega-p1}
\end{eqnarray}
where $[f(a)]^{a=p+1}_{a=1}\equiv f(p+1)-f(1)$, and $\kappa\equiv 2\mu_B^\ast
B$, $\mu_B^\ast\equiv e^\ast\hbar/2m^\ast c$ is the Bohr magneton of the
parafermion.
The contour is closed to the left
of the imaginary axis for
$\mu>0$, and thus
the sum of residues of (\ref{Omega-p1}) are evaluated
at the poles on both the negative real semiaxis $s=0,\,-1,\,-2,\,-3,\ldots$
(including $s=0$) and the imaginary axis $s_l=2\pi il k_B T/\kappa$.
The poles located on the real axis contribute to the ``monotonic part''
of $\Omega$, while the poles $s_l=2\pi il k_B T/\kappa$
dominate the behavior of its oscillating part.

We shall be interested only in the case of low temperature
in which $k_B T<\kappa$. Taking the sum at the poles
$s_l=2\pi il k_B T/\kappa$, and the pole $s=0$, we obtain $\Omega$
in the form
\begin{eqnarray}
\Omega &=& -n_2\kappa \left[p\left({z^2\over 8\pi^2}-\frac{1}{24}\right)
 + {5\pi^2\over 12}{p\over p+1}\left({k_B T\over \kappa}\right)^2\right]
 \nonumber \\
       & & -n_2k_B T  \sum_{l=1}^\infty {(-1)^{l+1} \over l}
\cos(2\pi l\mu/\kappa)
\left[\coth(2\pi^2 l k_B T/a\kappa)\right]^{a=p+1}_{a=1}~.
\label{Omega-p2}
\end{eqnarray}
The last term in (\ref{Omega-p2}) is the oscillatory part of $\Omega$.
We note here that in deriving the monotonic part
of $\Omega$, we took into account only the contribution of the
pole $s=0$, since the poles $s\ne 0$ contribute to terms of the order
$(-1)^{l+1}\exp\left\{-l\left(\mu+
\kappa/2\right)/k_B T\right\}/l$, which are exponentially small.
From (\ref{Omega-p2}) one can evaluate the mean particle number
$N=-\partial\Omega/\partial\mu$
and the magnetisation ${\bf M}=-\partial\Omega/ \partial{\bf B}$
of the parafermion gas in the magnetic field at thermodynamic
equilibrium.

In the $T\to 0$ limit, (\ref{Omega-p2}) reduces to
\begin{eqnarray}
\Omega &=& -n_2\kappa \left[{z^2\over 8\pi^2}-\frac{1}{24}
 + {1\over 2\pi^2} F(z)\right]\cdot p~,\ \
\nonumber \\
F(z) & =&  {\pi^2\over 12}-{1\over 4}z^2~~~,~~ -\pi\leq z \leq \pi~.
\label{Omega-p3}
\end{eqnarray}
where  $z\equiv 2\pi\mu/\kappa$.
From (\ref{Omega-p3}), we obtain the mean particle number
of the gas to be:
\begin{eqnarray}
  \frac{N(z)}{n_2}&=&{1\over 2\pi}\left(z + 2~{dF(z)\over dz}\right)\cdot p~
\nonumber  \\
         &=&\cases{ 0~~,& $0\leq z < \pi~~$;\cr
                  i\cdot p~~,& $(2i-1)\pi < z < (2i+1)\pi~~$,\cr}
   \label{N-p} \\
         & & i=1,2,3,\ldots \nonumber
\end{eqnarray}
Eq.(\ref{N-p}) gives the mean particle number at equilibrium as a function of
$\mu$ and $B$.  One sees that $N/n_2$ assumes values which are positive
integral multiples of $p$, as long as the chemical potential $\mu$ lies
between two Landau levels.
It is, however, the number of particle $N$ of the gas that is usually
assumed
given.  Eq.~(\ref{N-p}) then gives $\mu$ as an implicit function of $N$ and
$B$. Inverting (\ref{N-p}), one has, for $T\to 0$:
\begin{eqnarray}
\frac{\mu}{\kappa}
&=& i+\frac{1}{2}~~, ~~~~~ip< N/n_2 <(i+1)p~~,
\label{mu-p} \\
& &~~~i=0,1,2,\ldots         \nonumber
\end{eqnarray}

To obtain the magnetisation $M_z$ ($M_x=M_y=0$) at $T=0$, it is not useful to
take the derivative
of the $\Omega$ in (\ref{Omega-p3}) with respective to the magnetic field $B$.
This is because one must eliminate the chemical potential $\mu$ in the
resulting expression using (\ref{mu-p}).  But the function
$F(z)$ is singular at these values of $\mu/\kappa$.
One can, however, follow exactly the steps in \cite{Huang} by evaluating first
the total energy $U$ of the system.  Suppose the total number of parafermions
is $N$, and the field $B$ is such that the $i$ lowest Landau levels are
completely filled, and the $(i+1)$-th level is partially filled. The internal
energy of the gas is
\begin{eqnarray}
U(B) = \cases{\mu_B^\ast BN~~, & $0 < N/n_2 < p$~~;\cr
                \mu_B^\ast BN\left[(2i+3) - (i+1)(i+2)pn_2/N\right]~~,
                     & $(i+1)p < N/n_2 < (i+2)p$~~.\cr  }
\label{U-p}
\end{eqnarray}
The magnetisation is then given by $M_z=-\partial U/\partial B$:
\begin{eqnarray}
M_z(B) = \cases{-\mu_B^\ast N~~, & $0 < N/n_2 < p$~~;\cr
                \mu_B^\ast N\left[2 (i+1)(i+2)pn_2/N-(2i+3)\right]~~,
                     & $(i+1)p < N/n_2 < (i+2)p$~~.\cr  }
\label{M-p}
\end{eqnarray}
The magnetic susceptibility is obtained by taking the $B$-derivative of
(\ref{M-p}).
Eq.~(\ref{M-p}) shows that $M_z$ oscillates with period $p$ as the filling
factor $N/n_2$ changes.  This is the Landau-de Haas-van Alphen effect
of the 2D parafermions.  Setting $p=1$ gives the results for the fermions
\cite{Huang}.

\section{2D $g$-on gas}

We now consider the case of a 2D $g$-on gas in a uniform magnetic field $B$
along the $z$ direction.
The energy spectrum is again the Landau levels
\begin{eqnarray}
E_j=\kappa\left(j+{1\over 2}\right)~,~~~~
j=0,1,2,\ldots
\label{f2}
\end{eqnarray}
where
$\kappa=2\mu_B^\ast B$ as
in Section 2, but with
$m^\ast$, $e^\ast$ and $\mu_B^\ast$ now refer to the effective mass, the
magnitude of the effective charge, and the Bohr magneton of the $g$-on,
respectively.

The thermodynamical potential of $g$-on gas is given by \cite{Wu}
\begin{eqnarray}
\Omega &=&-k_B T n_2 \sum_{j=0}^{\infty}\ln\frac{1+(1-g){\bar
n}_j}{1-g{\bar n}_j}        \nonumber \\
       &=&-k_B T n_2 \sum_{j=0}^{\infty}\ln\frac{1+\omega_j}{\omega_j}~,
\label{TP-g}
\end{eqnarray}
where $n_2$ is the degeneracy of Landau levels as before, and $w_j$
is obtained by solving
(\ref{w}) with $\zeta_j=\exp\{[\kappa(j+1/2)-\mu]/k_B T\}$.  Use has
been made of (\ref{n-g}) in getting the second line in (\ref{TP-g}).
To derive the expressions for the magnetisation $\bf M$ and
the mean number of particles $N$ at equilibrium, we need
the following expressions, which can be easily checked from (\ref{w}):
\begin{eqnarray}
{\partial \omega_j\over\partial \mu}&=& -\frac{1}{k_B T}
\frac{\omega_j  (\omega_j+1)}{\omega_j +g}~,
\nonumber \\
{\partial \omega_j\over\partial B}&=&
2\left(j+\frac{1}{2}\right)\frac{\mu_B^\ast}{k_B T}
\frac{\omega_j  (\omega_j+1)}{\omega_j +g}~.
\label{diff-w}
\end{eqnarray}
From (\ref{diff-w}) we obtain the mean particle number
\begin{eqnarray}
N=n_2\sum_j {\bar n}_j=n_2\sum_j \frac{1}{\omega_j+g}~,
\label{Num-g}
\end{eqnarray}
and the magnetisation of the $g$-on gas ($M_x=M_y=0$)
\begin{eqnarray}
M_z=\kappa\frac{e^\ast L_xL_y}{hc}\left[{k_B T\over
\kappa}\sum_j\ln\frac{1+\omega_j}{\omega_j} -\sum_j
\left(j+{1\over 2}\right)~{\bar n}_j\right]~.
\label{M-g}
\end{eqnarray}

It is difficult to evaluate eq.~(\ref{Num-g}) and (\ref{M-g}) analytically
as functions of $\mu,B$ and $T$.   We thus resort to
numerical method.  The results are depicted in Fig.~2 and Fig.~3.
In Fig.~2 we show plots of $2\mu/\kappa$ as a function of the filling
factor $N/n_2$ for $g=1/2$ and $g=2$ at $\kappa/k_B T=20$ and $200$. These
graphs shows that the chemical potential $\mu$ coincides with the Landau
levels at sufficiently low temperatures.  Fig.~3 presents plots of
$M_z/(\kappa e^\ast L_xL_y/hc)$
as a function of  $2\mu/\kappa$ for $g=1/3,2/3,1,2$ and $3$ at $\kappa/k_B
T=20$.
One clearly sees the oscillatory behaviour of this function at low
temperatures.
One must, however, be reminded that one has to eliminate the chemical
potential $\mu$ in (\ref{M-g}) by (\ref{Num-g}) in order to get the actual
value of the magnetisation.

We now consider the $T=0$ limit.  Eq.~(\ref{Num-g}) and (\ref{n-g-2}) gives
\begin{eqnarray}
\frac{N}{n_2}&=&\cases{ 0~~,& $0\leq \mu/\kappa < 1/2~~$;\cr
                  i/g~~,& $i-1/2 < \mu/\kappa< i+1/2~~$;\cr}
   \label{N-g} \\
         & & i=1,2,3,\ldots \nonumber
\end{eqnarray}
Inverting (\ref{N-g}), one obtains the chemical potential as a function of
the filling factor:
\begin{eqnarray}
\frac{\mu}{\kappa}
&=&i+\frac{1}{2}~~, ~~~i/g<N/n_2<(i+1)/g~~,
\label{mu-g} \\
& & i=0,1,2,\ldots         \nonumber
\end{eqnarray}
For $g=1/p$, these expressions reduce to (\ref{N-p}) and (\ref{mu-p}) for the
parafermions as expected.

Magnetisation of the 2D $g$-on gas at $T=0$ as a function of the filling factor
can be derived as in Section 2 following
the steps in \cite{Huang}.  Here we shall do it directly from the expression
of (\ref{M-g}) at $T=0$ limit.  First we note that as $T\to 0$, $\omega_j\to 0$
($\infty$) for $E_j<\mu$ ($E_j>\mu$).  By applying the L'Hospital rule to
the first term of (\ref{M-g}), and using the following equality (derivable
from (\ref{w}))
\begin{eqnarray}
{\partial \omega_j\over\partial \beta}=\left(E_j - \mu\right)
\frac{\omega_j  (\omega_j+1)}{\omega_j +g}~,~~~\beta\equiv 1/k_B T~,
\label{diff-w-T}
\end{eqnarray}
one gets the the expression of $M_z$ at $T=0$:
\begin{eqnarray}
M_z=\frac{2\mu_B^\ast n_2}{\kappa}\left[\sum_{E_j<\mu} {1\over g}
\left(\mu-E_j\right) - \sum_{E_j\leq\mu} {\bar n}_j E_j\right]~.
\label{M-g-2}
\end{eqnarray}
As in Section 2, we again suppose that the total number of $g$-ons
is $N$, and the field $B$ is such that the $i$ lowest Landau levels are
completely filled, and the $(i+1)$-th level is partially filled.
This implies that $\mu=E_{i+1}$.  In (\ref{M-g-2}) we must substitute
for ${\bar n}_j$ the values ${\bar n}_j=1/g$ for $j=0,1,2,\ldots,i$, and
${\bar n}_{i+1}=(N/n_2)-(i+1)/g$.  Simple algebraic manipulations then give
\begin{eqnarray}
M_z = \cases{-\mu_B^\ast N~~, & $0 < N/n_2 < 1/g$~~;\cr
                \mu_B^\ast N\left[2 (i+1)(i+2)(n_2/gN)-(2i+3)\right]~~,
                     & $(i+1)/g < N/n_2 < (i+2)/g$~~.\cr  }
\label{M-g-3}
\end{eqnarray}
Eq.~(\ref{M-g-3}) shows the Landau-de Haas-van Alphen effect of the 2D
$g$-ons. It is identical with (\ref{M-p}) for $g=1/p$, and reduces to the
expression for the fermions at $g=1$.

\section{Qumtum Hall effect}

Let us envisage a hypothetical situation in which the
charge
carriers in the usual integral quantum Hall effect (IQHE)  were taken to be the
$g$-ons or the parafermions, instead of the electrons.
Let us take $g$-ons for definiteness.
The case with parafermions can be considered in the same way.
As in the IQHE, we assume that there is disorder potential
which broadens the Landau levels, giving rise to localized and extended states.
The chemical potential is assumed to lie between the Landau levels as a
consequence of the presence of the localized states.
Only the extended states, which coincide with the centres of the Landau levels,
contribute to the
Hall conductance.  For very weak external electric fields,
the mean number of $g$-ons in the extended states is well approximated
by (\ref{Num-g}) at low $T$, and by (\ref{N-g}) at $T=0$.
We show in Fig.~4 $N/n_2$ as a function of
$2\mu/\kappa$ (at fixed $\mu$ and $T$) for $g=1/3,2/3, 3, 1/5$ and $2/5$  at
various values of
$\mu/k_B T$.
It is obvious that $N/n_2$ exhibits plateaux with values  $i/g$ at low
temperatures, where $i$ equals the integral part of ($\mu/\kappa + 1/2$).
The Hall conductivity can be expressed as:
\begin{eqnarray}
  \sigma_{yx}={e^\ast cN\over BL_xL_y}\equiv{e^2\over h}\cdot
\left(e^\ast\over e\right)^2\cdot {N\over n_2}.
\label{hall-1}
\end{eqnarray}
Consequently, from
(\ref{hall-1}) and (\ref{N-g}), we see that the Hall conductance $\sigma_{yx}$
as a function
of $\mu/B$ is quantized at:
\begin{eqnarray}
  \frac{\sigma_{yx}}{e^2/h}=\frac{i}{g}\cdot (e^\ast/e)^2~.
\label{hall-2}
\end{eqnarray}
Thus depending on the values of $e^\ast$ and $g$, the IQHE of $g$-ons in
general gives a FQHE in the usual sense.

Consider now $g$-ons with parameters $g\equiv
g_{\mp}(m,p)=1\mp 2p/(2pm+1)$ and charge $\pm e^\ast (m,p)=\pm 1/(2pm+1)$ ($p$
and $m$ being integers).
Such parameters are relevant to the quasiholes (quasielectrons) in Jain's
composite fermion approach to the FQHE with ``filling factor"  (as
defined in the FQHE)
$\nu\equiv \sigma_{yx}/(e^2/h)=m/(2pm+1)$ (\cite{Isa1},
\cite{Shirai}).
For a 2D $g$-on gas with these parameters, one obtains from (\ref{hall-2})
\begin{eqnarray}
  \frac{\sigma^{(\mp)}_{yx}}{e^2/h}=\frac{i}{(2pm+1)[2p(m\mp 1)+1]}~.
\label{hall-qhe}
\end{eqnarray}
It should be noted that the observed filling factor $\nu=m/(2pm+1)$
in FQHE
occurs in our case only for $i=m\cdot [2p(m\mp 1)+1]$ for $g$-ons
with $g=g_{\mp} (m,p)$ and charge $\pm e^\ast (m,p)$.
The form of (\ref{hall-qhe}) also suggests that
$g$-ons with $g_{+}(m,p)$ and charge $-e^\ast(m,p)$ are equivalent to
$g$-ons with $g_{-}(m+1,p)$ and charge $+e^\ast(m+1,p)$, in that they give the
same Hall conductance.  This duality between the two kinds of $g$-ons is first
discussed in \cite{Shirai} from a different consideration.

\section{Summary}

We have studied the low temperature behaviours of 2D ideal
gases of generalized statistics in a uniform magnetic field.  The generalized
statistics considered here are the parafermion statistics and the exclusion
statistics. The magnetisation and the mean particle number of the gases are
obtained.
One notices that the properties of the parafermion gas of order $p$
are very similar to those of the $g$-on gas with $g=1/p$ at low temperatures.
These two systems become exactly equivalent at $T=0$.
We also briefly discuss quantum Hall effect with these particles as charge
carriers.

\newpage
\centerline{\bf Acknowlegdment}

This work is supported in part by the R.O.C. Grants number
NSC-86-2112-M-032-002.

\newpage

\centerline{\bf Figures Captions}
\vskip 2 truecm

Fig.~1.  Plots of statistical distribution ${\bar n}_i$ versus $E_i/\mu$
         for the exclusion (solid curves) and parafermion (dashed curves)
         statistics at $\mu/k_B T=15$.
\vskip 1 truecm

Fig.~2.  Plots of $2\mu/\kappa$ as a function of the filling factor
         $N/n_2$ for the exclusion statistics with $g=1/2$ and $2$ at
         $\kappa/k_B T=20$ (dashed curves) and $200$ (solid curves).
\vskip 1 truecm

Fig.~3.  Plots of $M_z/(\kappa e^\ast L_x L_y/hc)$ as a function of
         $2\mu/\kappa$ for the exclusion statistics at $\kappa/k_B T=20$.
\vskip 1 truecm

Fig.~4.  Plots of $N/n_2$ as a function of $2\mu/\kappa$ for the
         exculsion statistics with : (a) $g=1/3,2/3$ and $3$ at $\mu/k_B T=10$
        (dotted curves), $30$ (dashed curves) and $150$ (solid curves); (b) $g=
         1/5$ and $2/5$ at $\mu/k_B T=100$.


\newpage

\begin{thebibliography}{2}

\bibitem{FS} For reviews on anyons, and their r\^oles in FQHE and anyon
superconductivity, see, for examples, papers collected in {\sl Quantum Hall
Effect}, ed. M. Stone (World Scientific, Singapore, 1992); {\sl Fractional
Statistics and Anyon Superconductivity}, ed. F. Wilczek
(World Scientific, Singapore, 1990); Y. Hosotani, {\sl Int. J. Mod. Phys.}
{\bf B7} (1993) 2219.
\bibitem{Haldane} F.D.M. Haldane, {\sl Phys. Rev. Lett.} {\bf 67} (1991) 937.
\bibitem{JohnCan} M.D. Johnson and G.S. Canright, {\sl Phys. Rev.}
{\bf B49} (1994) 2947.
\bibitem{Wu} Y.S. Wu, {\sl Phys. Rev. Lett.} {\bf 73} (1994) 922.
\bibitem{Su} W.P. Su, Y.S. Wu and J. Yang, {\sl Phys. Rev. Lett.} {\bf 77}
(1996) 3423.
\bibitem{Isa1} S.B. Isakov, G.S. Canright and M.D. Johnson, cond-mat/9608139.
\bibitem{Shirai} J. Shiraishi, M. Kohmoto and Y.S. Wu, cond-mat/9612017.
\bibitem{GG} G. Gentile, {\sl Nuovo Cimento} {\bf 17} (1940) 493; {\bf 19}
(1942) 109; H.S. Green, {\sl Phys. Rev.} {\bf 90} (1953) 270.
\bibitem{Gb} O.W. Greenberg, {\sl Phys. Rev. Lett.} {\bf 13} (1964) 598.
\bibitem{AVT} S.I. Ben-Abraham, {\sl Am. J. Phys.} {\bf 38} (1970) 1335;
M.C. de Sousa Vieira and C. Tsallis, {\sl J. Stat. Phys.} {\bf 48} (1987) 97.
\bibitem{NayWil} C. Nayak and F. Wilczek, {\sl Phys. Rev. Lett.} {\bf 73}
(1994) 2740.
\bibitem{Raja} A.K. Rajapokal, {\sl Phys. Rev. Lett.} {\bf 74} (1995) 1048.
\bibitem{Isa2} S.B. Isakov, D.P. Arovas, J. Myrheim and A.P. Polychronakos,
{\sl Phys. Lett.} {\bf A212} (1996) 299.
\bibitem{Yang} H.S. Yang, B.-H. Lee and C. Park, cond-mat/9606165.
\bibitem{Kha} V.R. Khalilov, {\sl Electrons in Strong Electromagnetic
  Fields: an advanced classical and quantum treatment} (Gordon \& Breach,
  Amsterdam, 1996).
\bibitem{Huang} K. Huang, {\sl Statistical Mechanics} (2nd. ed.) (John Wiley
\& Sons, Inc., New York, 1987), pp. 260-261.
\end{thebibliography}
\end{document}